\documentclass[twocolumn,showpacs,preprintnumbers]{revtex4}
\usepackage{amsfonts}
\usepackage{graphicx}
\usepackage{bm}

\input{tcilatex}

\begin{document}

\preprint{}
\title{Influence of a repump laser on a nearly degenerate four-wave-mixing \\
spectrum in atomic vapors}
\author{Wei Jiang}
\email{jwayne@mail.ustc.edu.cn}
\author{Qun-feng Chen}
\author{Yong-sheng Zhang}
\author{G. -C. Guo}
\affiliation{Key Laboratory of Quantum Information,\\
University of Science and Technology of China, Hefei, 230026,P. R. China}
\date{\today}

\begin{abstract}
The influence of a repump laser on a nearly degenerate four-wave-mixing
(NDFWM) spectrum was investigated. We found the amplitude and line shape of
the NDFWM depended strongly on the detuning of the repump field. A five-peak
structure was observed. And at some certain repump detuning a dip appeared
at the central peak. A rough analysis was proposed to explain this effect.
\end{abstract}

\maketitle

\section{INTRODUCTINON}

Nearly degenerate four-wave mixing (NDFWM) has been extensively studied
during last three decades, because it has many applications, such as phase
conjugating mirror and NDFWM spectroscopy \cite%
{Fisher,Liao,Bloch,Grynberg,Oria,JLiu,Zhu1,Lezama}. In the latter case
people are interested in the line shape of the spectrum. This spectrum can
provide important information about the relaxation of states due to
reservoir coupling. Many experiments employed alkali vapor as the nonlinear
medium. Various theoretical models were presented \cite{Boyd,Steel,Berman}.
These models were based on a two-level system (TLS) driven by a pump field.
However in the real case a pure TLS is hard to find, one must consider the
influence of Zeeman degeneracy and other effects. In this paper we report an
experimental study of NDFWM when a repump field is applied. We show that
this repump field not only increases the NDFWM signal dramatically but also
affects the lineshape of the NDFWM spectrum in a profound way. We present a
simple model to explain one single phenomenon. More efforts are still needed
to fully understand the phenomena that we observed.

\section{EXPERIMENT}

We used the D1 transition of $^{85}Rb$ to produce the NDFWM signal and a
repump laser was tuned to D2 line of $^{85}Rb$. Fig. 1 shows the energy
diagram of the atom and schematic setup of the experiment. The pump field
was detuned from $5S_{1/2}(F=2)\rightarrow 5P_{1/2}(F=3)$ transition. $%
\Delta $ is detuning of pump laser from this transition. The probe laser
scanned across this transition. The relative detuning of probe field to pump
field is $\delta .$ The repump laser was tuned near $5S_{1/2}(F=3)%
\rightarrow 5P_{3/2}$ manifold with detuning $\Delta _{r}$. $\Delta _{r}$ is
the detuning from $5S_{1/2}(F=3)\rightarrow 5P_{3/2}(F=2)$ transition. All
these lasers are external cavity diode laser (ECDL) and have a linewidth
below 1MHz. The powers of pump and probe fields are both 0.6mW. The power of
repump field is 3mW. In order to minimize the influence of Doppler
broadening, the angle between probe and pump beam were kept small (about
4mrad). Diameters of the probe and pump beam are about 1mm. After passing
the vapor cell the forward pump beam was reflected back to form the backward
pump beam and overlapped with the forward pump beam. The pump and probe
laser were both linearly polarized. The generated NDFWM signal was pick up
by a 50/50 beam splitter and directed to a photodiode detector. The
temperature of the vapor cell was about $60^{\circ }$C, which resulted in an
atomic density $3.5\times 10^{11}/cm^{3}$.

Fig. 2 shows the recorded signal without repump beam. The signal has a
triplet structure. This phenomenon has been reported in Ref. \cite{Zhu1},
but was not properly explained. In Ref. \cite{Zhu1} the three peaks were
said to be located at $\delta =0$ and $\delta =\pm \Omega ^{\prime },$where $%
\Omega ^{\prime }=\sqrt{\Delta ^{2}+\Omega ^{2}}$ is the generalized Rabi
frequency and $\Omega $ is Rabi frequency of the forward pump field. And a
picture based on dressed-state was given. However this is not correct.
Actually the two side peaks were located at ($\delta =\pm 2\Delta $)
respectively \cite{Steel}. Fig. 3 are positions of NDFWM peaks versus pump
detuning $\Delta $. We can see the experimental data and the theoretical
curve are in good agreement. Note that the produced NDFWM signal is rather
small. This is because of the optical pumping effect of the pump and probe
fields. These fields pumped most of the atoms to $5S_{1/2}(F=3)$ manifold
through optical pumping process. Atoms in this internal state will not
contribute to the NDFWM process, thus leads to a small signal. However when
we added a repump laser to pump these atoms back to the $5S_{1/2}(F=2)$
manifold, the produced NDFWM signal was dramatically changed. First the
signal increased significantly (Increased by a factor 20 when repump had an
appropriate detuning). This is easy to understand because the atoms which
can produce the NDFWM signal was increased due to hyperfine optical pumping
of the repump laser. It is more interesting that the lineshape of the NDFWM
signal changed dramatically too. We found this change depended sensitively
on the detuning of the repump laser. Fig. 4 shows the observed NDFWM signal
with a fixed pump detuning $\Delta \approx 105$MHz and various repump
detuning. The most remarkable change was that the original triplet structure
changed to a five-peak structure ($\Delta _{r}=-60$MHz). And at certain
repump detuning a dip appeared at the central peak. The relative amplitude
of these peaks also depended sensitively on the detuning of the repump
laser. The dip at the central peak appeared when the detuning of repump
field is about -79MHz. The FWHM of the dip was sub-Doppler and was dominated
by the residual Doppler width due to imperfect probe-pump aligning. As
pointed out in an early work by Berman \cite{Berman}, narrow structure was
expected to appear whenever the system did not conserve population,
orientation or alignment.

\section{DISCUSSIONS}

The five-peak structure is hard to explain although interesting. Here we
only give a possible explanation of these phenomena. The change of
amplitudes of these peaks maybe owing to the hole burning effect and the
energy level degeneracy. Because the linewidth of the repump laser is
narrow, it will burn a hole on the Maxwell velocity distribution of the
atoms. That is atoms with a certain velocity will experience the hyperfine
optical pumping effect. When the repump laser is scanned, atoms with
different velocity will be pumped back to $5S_{1/2}(F=2)$ manifold.
Consequently this will cause different lineshape of NDFWM signal at
different repump detuning $\Delta _{r}$. In the following we will try to
explain one single phenomenon that we observed. That is the dip in Fig. 4(c).

Consider a two level system as showed in Fig. 5. $\gamma _{1}$, $\gamma _{2}$
are total relaxation rates of level 1 and 2 respectively. $\gamma
_{2\rightarrow 1}$ are spontaneous decay from level 2 to level 1. $\Delta
=\omega -\omega _{0}$ and $\delta =\omega _{p}-\omega $ are detunings of the
pump field and the probe field respectively. We adopt the treatment in Ref.~%
\cite{Steel}. The equation which governs the evolution of density matrix $%
\rho $ is given by,%
\begin{widetext}
\begin{equation}
i\hbar(\frac{\partial}{\partial t}+\vec{v}\cdot\nabla)\rho
=[H_{0},\rho]+[V,\rho]-\frac{i\hbar}{2}[\Gamma,\rho]+i\hbar\frac{d\rho}%
{dt}|_{sp}+i\hbar\frac{d\rho}{dt}|_{ph}+i\hbar\Lambda\;,
\end{equation}
where $H_{0}$ is free Hamiltonian of the system, $V$ is the interaction term,
$\vec{v}\cdot\nabla$ accounts for the motion of the atoms. $\Gamma$
represents decay to the reservoir. $\frac{d\rho}{dt}|_{sp}$ and $\frac{d\rho
}{dt}|_{sp}$ describe decay from 2 to 1 and decay of coherence between them
respectively. $\Lambda$ is the term to account for incoherent pumping. The equations for the matrix elements are,%
\begin{eqnarray}
i\hbar(\frac{\partial}{\partial t}+\vec{v}\cdot\nabla)\rho
_{11}&=&(V_{12}\rho_{21}-c.c.)-i\hbar\gamma_{1}\rho_{11}+i\hbar\gamma
_{2\to1}\rho_{22}+i\hbar\lambda_{1}\;, \label{de1}
\\%
i\hbar(\frac{\partial}{\partial t}+\vec{v}\cdot\nabla)\rho
_{22}&=&-(V_{12}\rho_{21}-c.c.)-i\hbar\gamma_{2}\rho_{22}+i\hbar\lambda_{2}\;,\label{de2}
\\%
i\hbar(\frac{\partial}{\partial t}+\vec{v}\cdot\nabla)\rho
_{12}&=&-\hbar\omega_{0}\rho_{12}+(V_{12}\rho_{22}-\rho_{11}V_{21})-i\hbar
\gamma_{ph}^{T}\rho_{12}\;, \label{de3}%
\end{eqnarray}
where $\gamma_{ph}^{T}=\frac{1}{2}(\gamma_{1}+\gamma_{2})+\gamma_{ph}$.

The third order nonlinear polarization generated by $E_{f},$ $E_{b}$ and
$E_{p}$ is $P^{(3)}=\chi^{(3)}E_{f}E_{b}E_{p}^{*}$. The phase matching conditions
result in the signal field, $E_{s}$, counterpropagating with the probe
beam. If the pump fields are at frequency $\omega$ and the probe field at
frequency $\omega+\delta,$ then by energy conservation the frequency of the signal is $\omega-\delta$. Solving the density equations (\ref{de1}-\ref{de3}) in a perturbation manner can yield,%
\begin{eqnarray}
P^{(3)}&=&-\frac{N_{0}\mu_{12}}{4}\Omega_{f}\Omega_{b}\Omega_{p}^{\ast
}e^{-i[(\omega-\delta)t+\vec{k}_{p}\cdot\vec{r}]}%
\frac{1}{-(\Delta-\delta)-\vec k_{p}\cdot\vec
{v}+i\gamma_{ph}^{T}}\nonumber\\
&&\times[\frac{1-R}{\delta-\Delta\vec{k}%
\cdot\vec{v}+i\gamma_{1}}+\frac{1+R}{\delta-\Delta\vec
{k}\cdot\vec{v}+i\gamma_{2}}][\frac{1}{-\Delta+\vec k_{f}\cdot\vec{v}+i\gamma_{12}}+\frac{1}{(\delta+\Delta
)-\vec k_{p}\cdot\vec{v}+\gamma_{12}}]+c.c\;,
\end{eqnarray}
\end{widetext}where $\Omega _{i}\ (i=f,b,p)$ is the Rabi frequency $\mu
_{12}E_{i}/\hbar $ associated with optical field $E_{i}$. $N_{0}$ is the
equilibrium population difference,%
\begin{equation}
(\rho _{11}-\rho _{22})_{eq}=\frac{\lambda _{1}}{\gamma _{1}}-\frac{\lambda
_{2}}{\gamma _{2}}(1-\frac{\gamma _{2\rightarrow 1}}{\gamma _{1}})
\end{equation}

The spectrum lineshape is determined by the decay parameters mentioned
above. Two situations are of special interest in our case. Fig. 6(a) and (b)
shows the two NDFWM spectrum without integration over velocity distribution.
In Fig. 6(a), where $\gamma _{1}<\gamma _{2}-\gamma _{2\to 1},$ we can see
the triplet structure. A narrow peak in center and two broader peaks at the
wings. The linewidth of the center peak is determined by $\gamma _{1}$. When 
$\gamma _{1}>\gamma _{2}-\gamma _{2\to 1}$ the spectrum (Fig. 6(b)) is quite
different with the one shown in Fig. 6(a). While the essential difference is
that a dip appeared in the center of the spectrum. Fig. 6(c) and (d) are the
spectrums after integration over velocity. We can see the dip still exists
in Fig. 6(d), but the two side peaks are washed out by the integration.

Compare the phenomena we observed with the theory we can see that when a
repump laser with appropriate detuning is added the effective decay rate $%
\gamma _{2}-\gamma _{2\to 1}$ is dramatically reduced. Consequently a dip
will appear at the center of the spectrum. However quantitatively comparison
between experimental data and theoretical values is impossible because the
model we use is overly simplified. One must take the level degeneracy into
account. And the hole burning effect and power broadening of the repump
laser should also be taken into consideration. Therefore careful and complex
calculation is needed to achieve this goal.

We also changed the polarizations of pump, probe and repump beam
respectively. We found that the NDFWM signal was insensitive to these
changes. Only small variation of signal amplitude was observed.

In order to study the influence of the repump power on the NDFWM signal, we
did the experiment with different repump powers. We found when we increased
the repump power the signal increased while showed some saturation. When the
repump power was high enough, the dip in the center of the spectrum
disappeared. This is because the power broadening caused by the repump laser
washed out this tiny structure.

Finally we want to mention that we noticed there was a similar work reported
by Zhu \textit{et. al.} \cite{Zhu2}. But the phenomenon they observed was
totally different. In their work the D2 transitions of $^{85}Rb$ were used
to produce NDFWM signal and a repump laser was tuned to the D1 transitions.
They found no significant changes except the signal was amplified several
times. We conjecture that this is because their lasers were so strong that
saturation and power broadening became the dominant effects.

\section{CONCLUSION}

In conclusion, we have studied the influence of a repump laser on a NDFWM
spectrum. We found the amplitude and line shape of the NDFWM depended
strongly on the detuning of the repump field. A five-peak structure was
observed. And at a certain repump detuning a dip appeared at the central
peak. A rough analysis was proposed to explain this effect. More efforts are
still needed to fully understand the phenomena that we observed.

\begin{acknowledgments}
This work was funded by National Fundamental Research Program
(2001CB309300), National Natural Science Foundation of China (Grant No.
60121503, 10304017), the Innovation funds from Chinese Academy of Sciences.
\end{acknowledgments}

\vspace*{5ex}

\textbf{Figure Captions}

Fig. 1 Experimental Setup. P, polarizer; BS, 50/50 beam splitter; D, photo
diode detector. The angle between probe and pump beam were about 4mrad.
Diameters of the probe and the pump beam were about 1mm. After passing the
vapor cell the forward pump beam was reflected back to form the backward
pump beam and overlapped with the forward pump beam. The pump and probe
beams were 795nm lasers tuned near the $5S_{1/2}(F=2)\rightarrow
5P_{1/2}(F=3)$ transition of $^{85}Rb$ with detuning $\Delta $ and $\delta $
respectively. The repump beam was a 780 nm laser tuned near $5S_{1/2}(F=3)$ $%
\rightarrow $ $5P_{3/2}$ manifold with detuning $\Delta _{r}$. $\Delta _{r}$
is the detuning from $5S_{1/2}(F=3)\rightarrow 5P_{3/2}(F=2)$ transition.

Fig. 2 NDFWM spectrum without repump beam. The pump detuning $\Delta =115$%
MHz. $\delta $ is the relative detuning from the frequency of central peak.

Fig. 3 Peak positions of three NDFWM resonance versus pump detuning $\Delta. 
$

Fig. 4 NDFWM spectrum with repump field turned on and various repump
detuning $\Delta _{r}$. From (a) to (j) the repump detuning $\Delta _{r}$
are -205MHz, -132MHz, -79MHz, -60MHz, -15MHz, 93MHz, 122MHz, 163MHz, 168MHz
and 317 MHz respectively.

Fig. 5 Simple two-level system. $\gamma_{1}$ and $\gamma_{2}$ are total
decay rate of level 1 and 2 respectively. $\gamma_{2\to1}$ is the decay rate
from 2 to 1.

Fig. 6 Calculated NDFWM spectrum based on Eq.(5) with different parameters.
(a) $\Delta =50,\ \gamma _{1}=3,\ \gamma _{2}=6,\ \gamma _{ph}=3,\
\gamma_{2\to 1}=6$. (b) $\Delta =50,\ \gamma _{1}=3,\ \gamma _{2}=0.1,\
\gamma_{ph}=3,\ \gamma _{2\to 1}=6$. (c) the same as (a) but the spectrum
was integrated over velocity distribution. (d) the same as (b) but the
spectrum was integrated over velocity distribution.

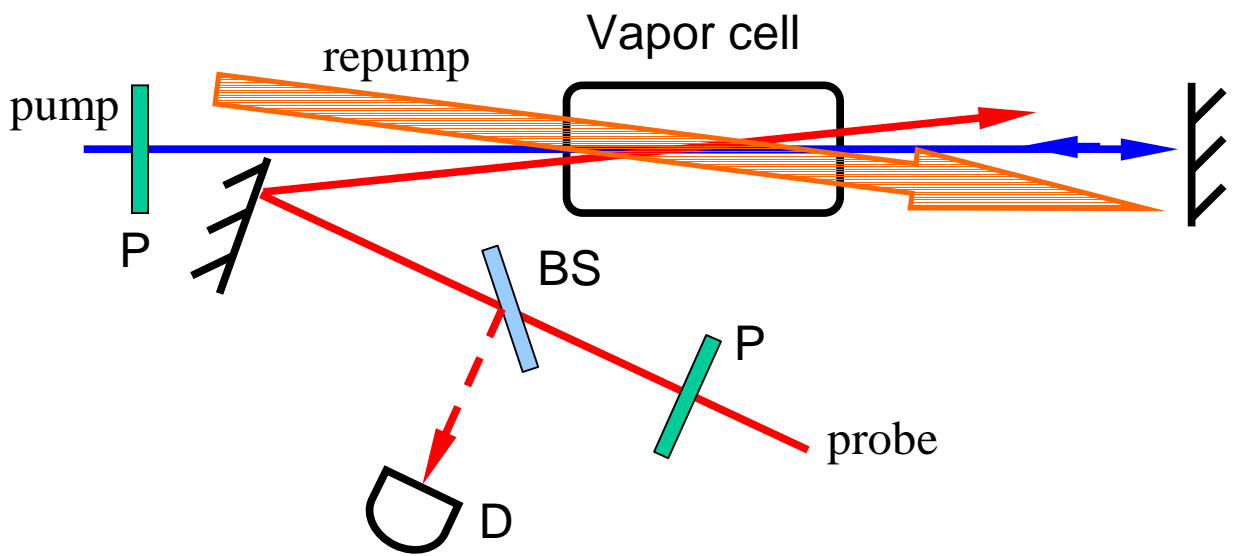
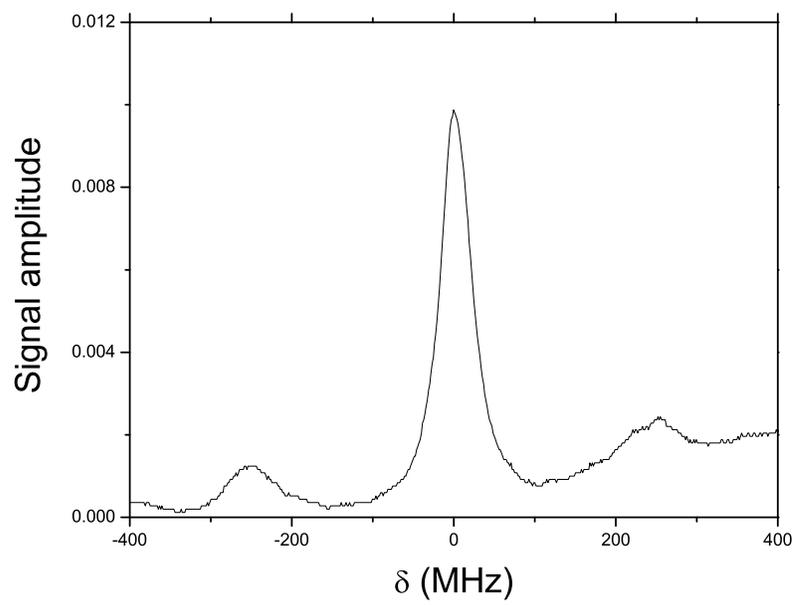
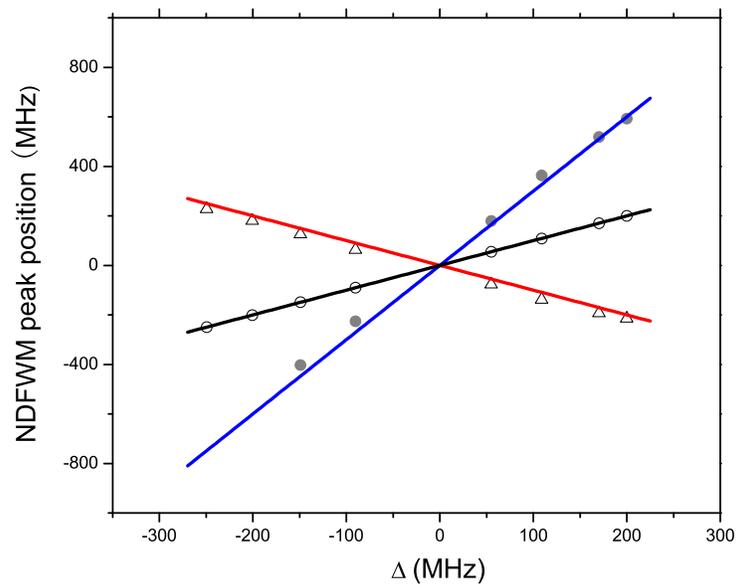
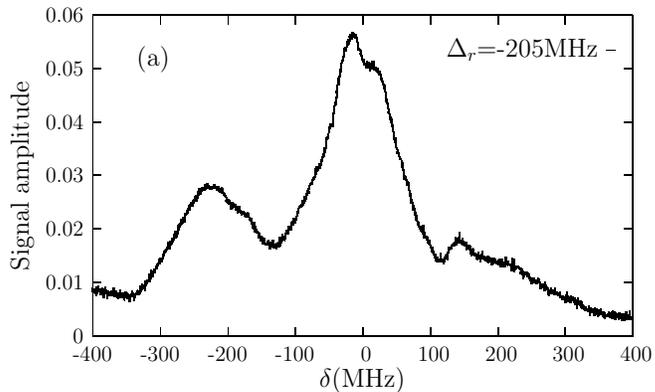
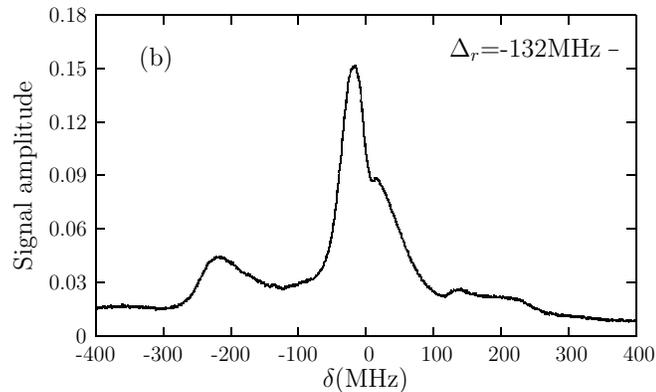
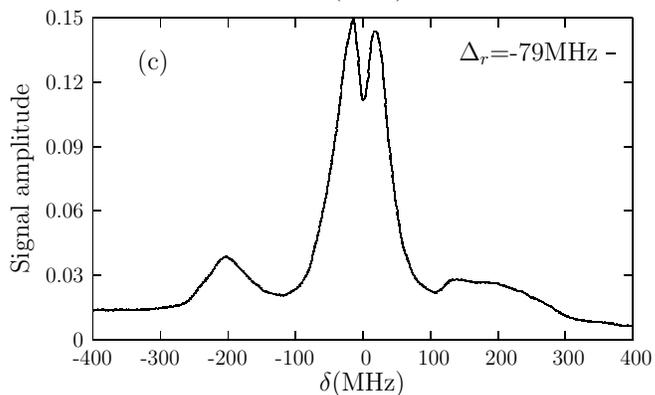
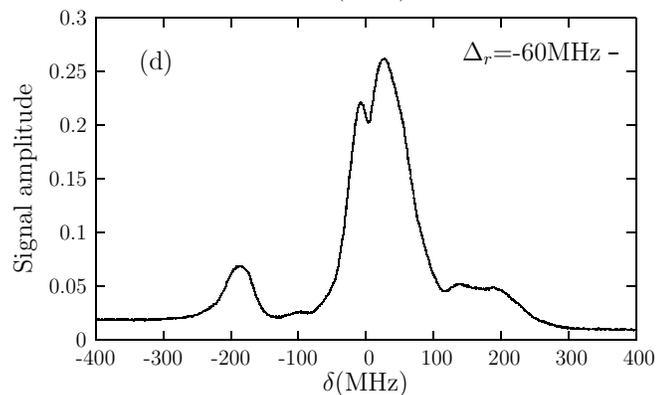
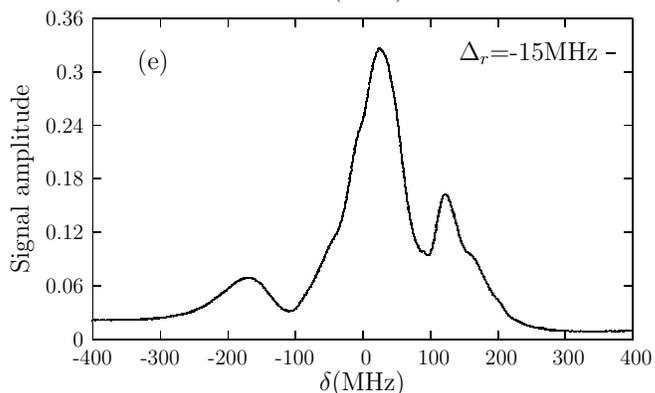
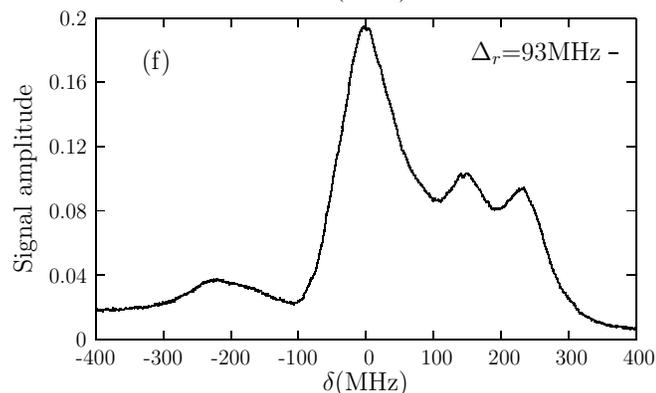
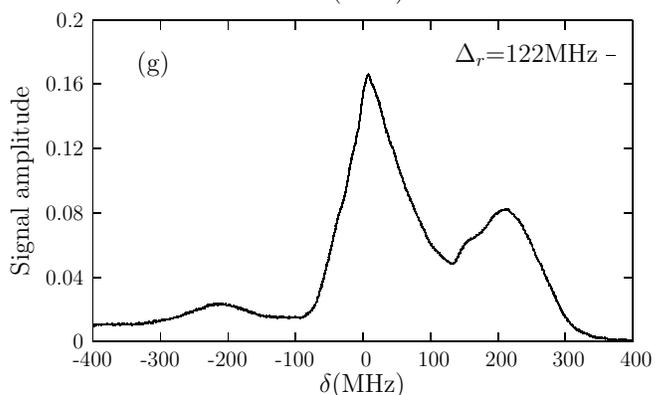
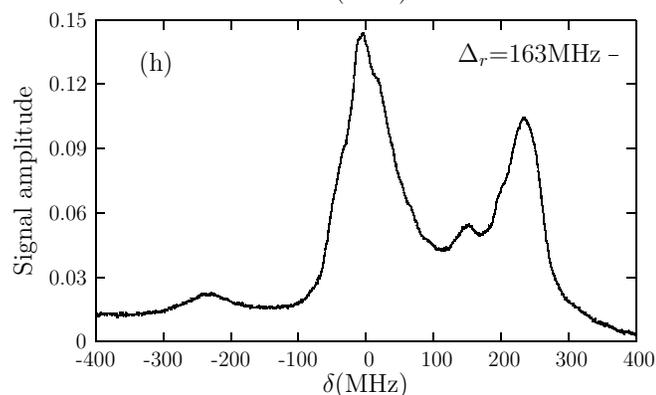
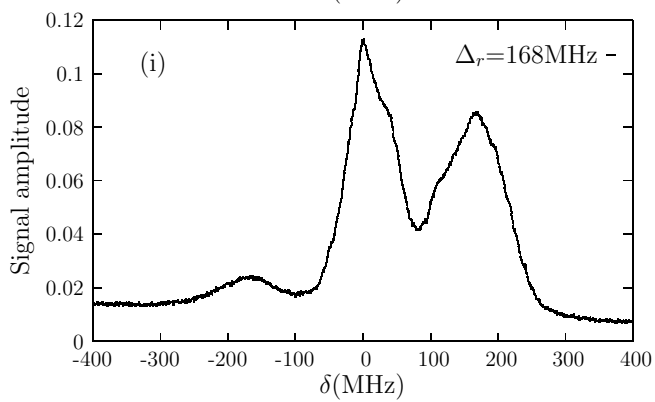
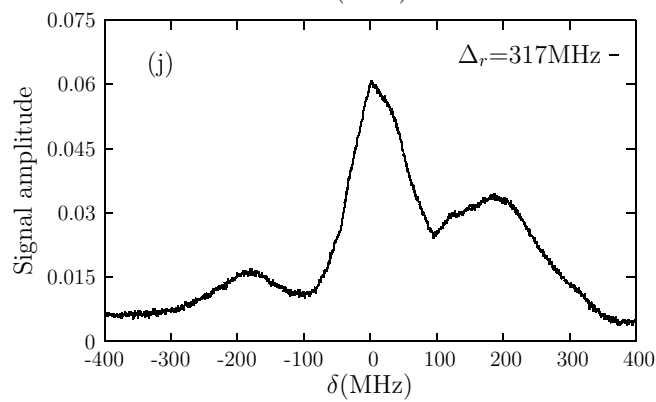
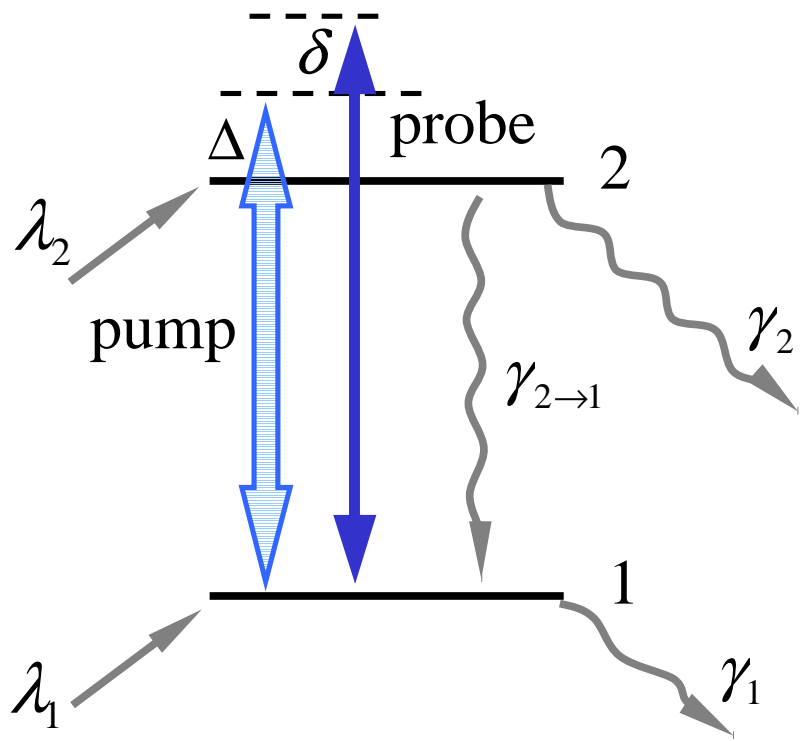
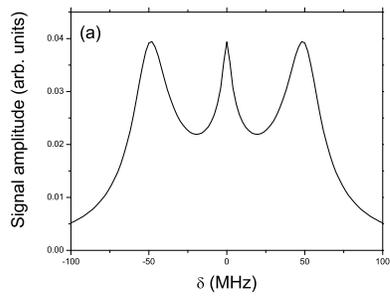
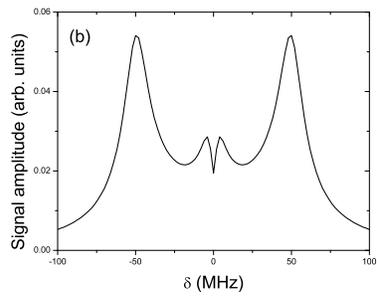
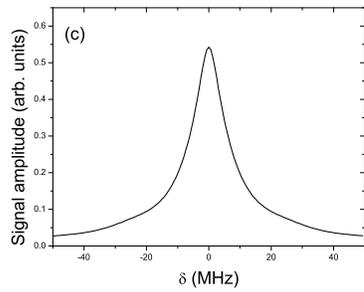
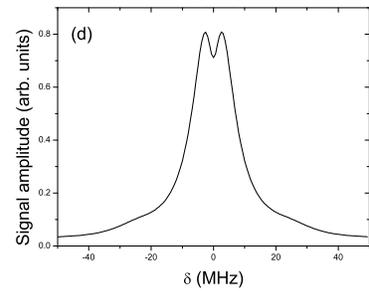
\end{document}